\documentclass[12pt]{article}
\usepackage{amssymb}
\usepackage{amsmath}
\def\d{\delta}

\def\la{\lambda}

\def\be{\begin{equation}}
\def\ee{\end{equation}}
\def\arr{\begin{array}{rll}}
\def\ea{\end{array}}
\def\bea{\begin{eqnarray}}
\def\eea{\end{eqnarray}}

\def\N2{$N{=}2$}

\def\>{\rangle}
\def\<{\langle}
\def\+{\dagger}
\def\={\ =\ }

\begin{document}
\renewcommand{\thefootnote}{\arabic{footnote}}
\noindent
\begin{titlepage}
\setcounter{page}{0}
\begin{flushright}
$\qquad$
\end{flushright}
\vskip 3cm
\begin{center}
{\LARGE\bf{The odd-order Pais-Uhlenbeck oscillator}}
\vskip 1cm
$
\textrm{\Large Ivan Masterov\ }
$
\vskip 0.7cm
{\it
Laboratory of Mathematical Physics, Tomsk Polytechnic University, \\
634050 Tomsk, Lenin Ave. 30, Russian Federation}
\vskip 0.7cm
{E-mail: masterov@tpu.ru}

\end{center}
\vskip 1cm
\begin{abstract}
\noindent
We consider a Hamiltonian formulation of the $(2n+1)$-order generalization of the Pais-Uhlenbeck oscillator with distinct frequencies of oscillation.
This system is invariant under time translations. However, the corresponding Noether integral of motion is unbounded from below and can be presented as a direct sum of $2n$ one-dimensional harmonic oscillators with an alternating sign. If this integral of motion plays a role of a Hamiltonian, a quantum theory of the Pais-Uhlenbeck oscillator faces a ghost problem. We construct an alternative canonical formulation for the system under study to avoid this nasty feature.
\end{abstract}

\vskip 1cm
\noindent
PACS numbers: 11.30.-j, 11.25.Hf, 02.20.Sv

\vskip 0.5cm

\noindent
Keywords: Pais-Uhlenbeck oscillator, ghost problem

\end{titlepage}

\noindent
{\bf{\large 1. Introduction}}
\vskip 0.5cm

The Pais-Uhlenbeck (PU) oscillator \cite{Pais} is one of the simplest higher-derivative mechanical models. In general, one constructs a Hamiltonian formulation of this system with the aid of Ostrogradsky's method \cite{Ostrogradski}. But the Hamiltonian obtained in such a way is unbounded from below. This leads to a ghost problem on quantization \cite{Pais,Woodard,Smilga}. The research of higher-derivative theories of gravity stimulates considerable interest in solving this long-standing problem. To obtain a more physically viable quantum theory of the PU oscillator, the efforts have been focused mostly on the construction of alternative Hamiltonian formulations and quantization procedures \cite{Kosinski}-\cite{Alt}. In particular, in \cite{Kosinski} it has been shown that a canonical formulation of the fourth-order PU oscillator is not unique. Moreover, in contrast to Ostrogradsky's approach, alternative canonical formalism may correspond to a positive-definite Hamiltonian.

According to the analysis in \cite{Pais} (see also \cite{Smilga}), the Ostrogradsky's Hamiltonian of the $2n$-order PU oscillator with distinct frequencies of oscillation can be presented as a direct sum of $n$ decoupled harmonic oscillators with an alternating sign. This provides a set of $n$ functionally independent positive-definite integrals of motion. The method introduced in \cite{Kosinski} is based on observation that a linear combination of these integrals with arbitrary nonzero coefficients can play a role of a Hamiltonian for the case of the fourth-order PU oscillator. So, we have a positive-definite Hamiltonian when all coefficients are positive. The corresponding Poisson bracket can be obtained as a solution of a nondegenerate system of linear equations. The same approach allows to construct an alternative canonical formulation for the PU oscillator of an arbitrary even order with distinct frequencies of oscillation \cite{Alt}.

In recent time the PU oscillator has also attracted much attention within the context of dynamical realizations of nonrelativistic conformal groups \cite{KB}, \cite{Galajinsky_1}-\cite{inf}. In particular, it has been shown that the $2n$-order PU oscillator for a particular choice of its frequencies of oscillation enjoys $l=n-\frac{1}{2}$-conformal Newton-Hooke symmetry \cite{Henkel}-\cite{Galajinsky_3}. On the other hand, the analogous dynamical realization of the $l$-conformal Newton-Hooke algebra for integer $l=n$ represents a higher-derivative model which is naturally called as $(2n+1)$-order PU oscillator \cite{Masterov_2,Masterov_3,Masterov_1}. Some aspects of the third-order PU oscillator have been studied in papers \cite{Lukier_1}-\cite{Liu} (see also \cite{Horvathy_1}-\cite{Gomis_1}). But the odd-order PU oscillator for values of order higher than three remains completely unexplored. At the same time, a construction of a Hamiltonian formulation of this model is an important issue for further possible quantum mechanical applications. The purpose of this work is to generalize the analysis obtained in papers \cite{Kosinski,Alt} to the case of the PU oscillator of the arbitrary odd order with distinct frequencies of oscillation.

The paper is organized as follows. In the next section we give the notion of the odd-order PU oscillator. In Sect. 3, we construct an alternative Hamiltonian formulation for the third-order PU oscillator.  A Hamiltonian formulation for the PU oscillator of the arbitrary odd order is considered in Sect. 4. Sect. 5 is devoted to possible generalizations of the odd-order PU oscillator which are compatible with alternative canonical formalism. In the concluding Sect. 6, we summarize our results and discuss further possible developments. Some technical details are gathered in Appendix A.

\vskip 0.5cm
\noindent
{\bf{\large 2. The model}}
\vskip 0.5cm

The recent results on dynamical realizations of so-called $l$-conformal Galilei algebra \cite{Henkel}-\cite{Negro_2} have shown that such a model as a free $(2l+1)$-order derivative particle exhibits this symmetry \cite{Gomis} (see also \cite{Horvathy_2}-\cite{Andr_6})\footnote{About realizations of $l$-conformal Galilei algebra without higher derivatives see \cite{Galajinsky_1}, \cite{AB}, \cite{WHD_1,WHD_2}.}. The action functional of this model for half-integer $l$ has a form\footnote{The summation over repeated spatial indices is understood, unless otherwise is explicitly stated.}
\bea\label{half}
S=\frac{1}{2}\int\,dt\,x_i x_i^{(2l+1)},
\eea
where $i=1,2,..,dim\,V$ is a spatial index; a superscript in braces designates the number of derivatives with respect to time. While for integer $l$ one has
\bea\label{int}
S=\frac{1}{2}\int\,dt\,\epsilon_{ij}\,x_i x_j^{(2l+1)},
\eea
where $\epsilon_{ij}$ is the Levi-Civit\'{a} symbol with $\epsilon_{12}=1$; $i,j=1,2$.

In papers \cite{PU,Niederer_1} (see also \cite{Andrzejewski_1}) it is shown that applying appropriate coordinate transformations to the models (\ref{half}), (\ref{int}) leads to counterparts of these systems in Newton-Hooke spacetime with a negative cosmological constant, i.e. in nonrelativistic spacetime with universal cosmological attraction (see e.g. \cite{Horvathy_2}, \cite{Galajinsky_6}-\cite{Galajinsky_4} and references therein). Newton-Hooke counterpart of the model (\ref{half}) is described by the action functional
\bea\label{halfNH}
S=\frac{1}{2}\int\,dt\,x_i\prod_{k=0}^{l-\frac{1}{2}}\left(\frac{d^2}{dt^2}+\frac{(2k+1)^2}{R^2}\right)x_i,
\eea
where $\Lambda=-\frac{1}{R^2}$ is a nonrelativistic cosmological constant. This action describes the even-order PU oscillator with a particular choice of its frequencies of oscillation.

The analogue of the action (\ref{halfNH}) for the integer $l=n$ is given by \cite{Masterov_2,Masterov_3,Masterov_1}\footnote{In papers \cite{Lukier_1,Lukier_2,Liu} the third-order PU oscillator has been obtained as a dynamical realization of Galilei and Newton-Hooke algebras.}
\bea\label{odd_conf}
S=\frac{1}{2}\int\,dt\,\epsilon_{ij}\,x_i\prod_{k=1}^{n}\left(\frac{d^2}{dt^2}+\frac{(2k)^2}{R^2}\right)\frac{dx_j}{dt},
\eea
which is naturally understood as corresponding to a particular case of the $(2n+1)$-order PU oscillator. So, the general form of the action functional of the odd-order PU oscillator reads
\bea\label{odd}
S=\frac{1}{2}\int\,dt\,\epsilon_{ij}\,x_i\prod_{k=0}^{n-1}\left(\frac{d^2}{dt^2}+\omega_k^2\right)\frac{dx_j}{dt}.
\eea
In the present paper we consider only the case when all frequencies of oscillation $\omega_k$, $k=0,1,..,n-1$ are distinct and nonzero.

\vskip 0.5cm
\noindent
{\bf{\large 3. The third-order PU oscillator}}
\vskip 0.5cm
Let us consider the action functional of the third-order PU oscillator
\bea\label{PU3}
S=\frac{1}{2}\int\,dt\,\epsilon_{ij}\,x_i\left(\frac{d^2}{dt^2}+\omega_0^2\right)\frac{d x_j}{dt}.
\eea
The dynamics of this model obeys the following equations of motion
\bea\label{3}
\dddot{x}_i+\omega_0^2\dot{x}_i=0,
\eea
where the dot denotes the derivative with respect to time.

In general, one constructs a canonical formulation for a higher-derivative theory by using of such approaches as Ostrogradsky's method \cite{Ostrogradski}, Dirac's method for constrained systems, or Faddeev-Jackiw's prescription \cite{Faddeev}. Let us obtain a Hamiltonian formulation of the model (\ref{PU3}) by applying Dirac's method. To this end we introduce the following first-order equivalent of the action (\ref{PU3})
\bea
S=\int dt\left[-\frac{1}{2}\epsilon_{ij}\left(y_i\dot{y}_j-\omega_0^2 x_i y_j\right)+\lambda_i(y_i-\dot{x}_i)\right],
\nonumber
\eea
where $y_i$ are new variables, $\lambda_i$ are Lagrange multipliers. Canonical momenta $p_i^x$, $p_i^y$, and $p_i^{\lambda}$ which correspond to $x_i$, $y_i$, and $\lambda_i$, respectively, are defined by a standard way
\bea
p_i^x=\frac{\partial L}{\partial\dot{x}_i}=-\lambda_i,\quad p_i^y=\frac{\partial L}{\partial\dot{y}_i}=\frac{1}{2}\epsilon_{ij}y_j, \quad p_i^{\lambda}=\frac{\partial L}{\partial\dot{\lambda}_i}=0.
\nonumber
\eea
So, canonical formalism of the model (\ref{PU3}) can be formulated in the twelve-dimensional phase space $(x_i,y_i,\lambda_i,p_i^x,p_i^y,p_i^\lambda)$ with six constraints
\bea
\phi_i^x=p_i^x+\lambda_i,\quad \phi_i^y=p_i^y-\frac{1}{2}\epsilon_{ij}y_j,\quad \phi_i^{\lambda}=p_i^{\lambda},\qquad i,j=1,2.
\nonumber
\eea
Therefore, the phase space of the third-order PU oscillator is six-dimensional. We can choose $(x_i,p_i^{x},y_i)$ as independent canonical variables which obey the following nonvanishing relations under corresponding Dirac bracket
\bea\label{DB}
\{x_i,p_j^x\}=\d_{ij},\qquad \{y_i,y_j\}=\epsilon_{ij}.
\eea
The Hamiltonian of the model (\ref{PU3}) is found to be
\bea\label{DH}
H=p_i^x y_i-\frac{\omega_0^2}{2}\epsilon_{ij}x_i y_j.
\eea
The analogues Hamiltonian formulation for the third-order PU oscillator with imaginary frequency of oscillation has been derived in \cite{Liu} with the aid of Faddeev-Jackiw prescription \cite{Faddeev} (see also \cite{Motohashi}).

The dynamics of phase space variables obey the following equations
\bea\label{DE}
\dot{x}_i=y_i,\qquad \dot{p}_i^x=\frac{\omega_0^2}{2}\epsilon_{ij}y_j,\qquad \dot{y}_i=\epsilon_{ij}p_j^x-\frac{\omega_0^2}{2}x_i.
\eea
It is easy to check that this system of equations is equivalent to (\ref{3}). The equations (\ref{DE}) allow to obtain a following representation of the Hamiltonian (\ref{DH}) in terms of variables $\dot{x}_i$, $\ddot{x}_i$
\bea\label{energy}
H=-\epsilon_{ij}\dot{x}_i\ddot{x}_j.
\eea
It is Noether integral of motion of the third-order PU oscillator which corresponds to invariance under time translations, as it should be.

It can be also straightforwardly verified that the following relations
\bea
&&\label{Eq3}
\{x_i,H\}=\dot{x}_i,\qquad\quad \{\dot{x}_i,H\}=\ddot{x}_i,\qquad\, \{\ddot{x}_i,H\}=-\omega_0^2\dot{x}_i;
\\[2pt]
&&\label{Br3}
\{x_i,\ddot{x}_j\}=-\epsilon_{ij},\qquad \{\dot{x}_i,\dot{x}_j\}=\epsilon_{ij},\qquad \{\ddot{x}_i,\ddot{x}_j\}=\omega_0^2\epsilon_{ij},
\eea
hold with regard to (\ref{DB}).

Let us introduce the variables
\bea\label{coord3}
\begin{aligned}
&
q_i=\sqrt{\frac{1}{2\omega_0}}\left(\dot{x}_1+\frac{(-1)^{i}}{\omega_0}\ddot{x}_2\right),\quad p_i=\sqrt{\frac{\omega_0}{2}}\left(\dot{x}_2+\frac{(-1)^{i+1}}{\omega_0}\ddot{x}_1\right),
\\[2pt]
&
\qquad\qquad\qquad\quad z_i=\frac{(-1)^i}{\omega_0}(\ddot{x}_i+\omega_0^2 x_i),\quad\mbox{(no sum)},
\end{aligned}
\eea
which obey the relations $\{q_i,p_j\}=\d_{ij}$, $\{z_i,z_j\}=\epsilon_{ij}$ under (\ref{Br3}). It is evident that these variables can be obtained by canonical transformation of coordinates $(x_i,y_i,p_i^x)$. It should be noted that expressions (\ref{coord3}) do not provide a finite limit $\omega_0\rightarrow 0$.

The Hamiltonian (\ref{energy}) in terms of variables (\ref{coord3}) takes a form
\bea\label{Ham3osc}
H=\frac{1}{2}(p_1^2+\omega_0^2 q_1^2)-\frac{1}{2}(p_2^2+\omega_0^2 q_2^2).
\eea
So, the Hamiltonian of the third-order PU oscillator can be presented as a direct sum of harmonic oscillators which alternate in a sign. An analogous representation of the Hamiltonian of the even-order PU oscillator has been obtained in the original paper \cite{Pais}. At first sight it may appear that the third-order PU oscillator is dynamically equivalent to a system of two decoupled one-dimensional harmonic oscillators. It is not true because the phase spaces of these two systems are not isomorphic. In addition to oscillator degrees of freedom $(q_1,q_2,p_1,p_2)$, the phase space of the third-order PU oscillator involves coordinates $z_i$ whose dynamics obey the first-order differential equations. This also can be illustrated by rewriting the action functional (\ref{PU3}) in terms of variables $(q_i,\dot{q}_i,z_i,\dot{z}_i)$ (up to a total derivative term)
\bea
S=\frac{1}{2}\int dt\left[(\dot{q}_1^2-\omega_0^2 q_1^2)-(\dot{q}_2^2-\omega_0^2 q_2^2)-\epsilon_{ij}z_i\dot{z}_j\right].
\nonumber
\eea

The Hamiltonian (\ref{DH}) (or in the form (\ref{Ham3osc})) of the system (\ref{PU3}) derived by Dirac's method is unbounded from below\footnote{In paper \cite{Motohashi} it has been shown that a Hamiltonian of the odd-order derivative mechanical system of general type is unbounded from below.}. When conventional quantization scheme is applied, we face a ghost problem. The method to obtain an alternative Hamiltonian formulation has been introduced in \cite{Kosinski}. According to this approach we may deform both the Hamiltonian of the third-order PU oscillator (\ref{Ham3osc}) and Poisson structure (\ref{Br3}) in such a way that the equations (\ref{Eq3}) will be preserved.
Let us consider the following deformation of the Hamiltonian (\ref{Ham3osc})
\bea\label{H3alt}
\mathcal{H}=\frac{\gamma_1}{2}(p_1^2+\omega_0^2 q_1^2)+\frac{\gamma_2}{2}(p_2^2+\omega_0^2 q_2^2)= \frac{1}{2}(\gamma_1-\gamma_2)H+\frac{1}{4\omega_0}(\gamma_1+\gamma_2)(\ddot{x}_i^2+\omega_0^2\dot{x}_i^2),
\eea
where $\gamma_1$ and $\gamma_2$ are arbitrary nonzero constants. This integral of motion corresponds to symmetry transformations of the action functional (\ref{PU3})
\bea\label{timetr}
\d x_i=\{x_i,\mathcal{H}\}a=\left(\frac{1}{2}(\gamma_1-\gamma_2)\dot{x}_i-\frac{1}{2\omega_0}(\gamma_1+\gamma_2)\epsilon_{ij}\ddot{x}_j\right)a,
\eea
where $a$ is an infinitesimal parameter. These transformations represent pure time translations when $\gamma_1=-\gamma_2$. In this sense we can understand (\ref{timetr}) as a deformation of time translations.

Then we substitute an ansatz (\ref{H3alt}) into equations (\ref{Eq3}) instead of $H$. Using the standard properties of the Poisson bracket allows to obtain restrictions on Poisson structure relations between variables $x_i$, $\dot{x}_i$, $\ddot{x}_i$. Resolving these restrictions leads to the following structure:
\bea\label{PS3alt}
\begin{aligned}
&
\;\{x_i,\ddot{x}_j\}=-\frac{1}{2}\left(\frac{1}{\gamma_1}-\frac{1}{\gamma_2}\right)\epsilon_{ij},\;\,
\{\dot{x}_i,\dot{x}_j\}=\frac{1}{2}\left(\frac{1}{\gamma_1}-\frac{1}{\gamma_2}\right)\epsilon_{ij}, \;\,
\{\ddot{x}_i,\ddot{x}_j\}=\frac{\omega_0^2}{2}\left(\frac{1}{\gamma_1}-\frac{1}{\gamma_2}\right)\epsilon_{ij},
\\[7pt]
&
\qquad\qquad\quad\;\{x_i,\dot{x}_j\}=\frac{1}{2\omega_0}\left(\frac{1}{\gamma_1}+\frac{1}{\gamma_2}\right)\d_{ij},\qquad\;
 \{\dot{x}_i,\ddot{x}_j\}=\frac{\omega_0}{2}\left(\frac{1}{\gamma_1}+\frac{1}{\gamma_2}\right)\d_{ij}.
\end{aligned}
\eea
A few comments are in order. Firstly, if we put $\gamma_1 = 1$, $\gamma_2 = -1$ then we reproduce (\ref{Br3}). Secondly, the Poisson structure (\ref{PS3alt}) is degenerate when $\gamma_1=\gamma_2$. By this reason our consideration is restricted only for values $\gamma_1\neq\gamma_2$. Thirdly, different possible pairs $(\gamma_1,\gamma_2)$ correspond to canonically nonequivalent structures (\ref{PS3alt}), i.e. one structure cannot be obtained from another by canonical transformations (see related discussion in \cite{Kosinski}). So, we have constructed two-parametric family of Hamiltonian structures for the third-order PU oscillator.

Canonical coordinates which correspond to the Hamiltonian (\ref{H3alt}) are
\bea\label{qp}
\mathfrak{q}_i=\sqrt{|\gamma_i|}q_i,\quad
\mathfrak{p}_i=(-1)^{i+1}\mbox{sign}(\gamma_i)\sqrt{|\gamma_i|}p_i,\quad \mbox{(no sum)}\;\rightarrow\;\{\mathfrak{q}_i,\mathfrak{p}_j\}=\d_{ij},
\eea
where the sign$(x)$ is the standard signum function. With these variables, the Hamiltonian (\ref{H3alt}) may be written as
\bea
\mathcal{H}=\frac{\mbox{sign}(\gamma_1)}{2}(\mathfrak{p}_1^2+\omega_0^2 \mathfrak{q}_1^2)+\frac{\mbox{sign}(\gamma_2)}{2}(\mathfrak{p}_2^2+\omega_0^2 \mathfrak{q}_2^2).
\nonumber
\eea

The variables $z_i$ obey relations
\bea
\{z_i,z_j\}=\alpha_0^{-}\epsilon_{ij}=\frac{1}{2}\left(\frac{1}{\gamma_1}-\frac{1}{\gamma_2}\right)\epsilon_{ij}
\nonumber
\eea
with regard of the Poisson structure (\ref{PS3alt}).
These variables also can be redefined as follows
\bea\label{pi}
\pi_1=\frac{1}{\sqrt{|\alpha_0^{-}|}}z_1,\qquad \pi_2=\frac{\mbox{sign}(\alpha_0^{-})}{\sqrt{|\alpha_0^{-}|}}z_2\;\rightarrow\; \{\pi_i,\pi_j\}=\epsilon_{ij},\; \{\pi_i,\mathfrak{q}_j\}=\{\pi_i,\mathfrak{p}_j\}=0.
\eea
The existence of coordinates $(\mathfrak{q}_i,\mathfrak{p}_i,\pi_i)$ automatically proves that Jacoby identity for the Poisson bracket corresponding to (\ref{PS3alt}) holds.

So, if both coefficients $\gamma_1\neq\gamma_2$ are positive, we have the nondegenerate Poisson structure and positive-definite Hamiltonian. This leads to the ghost-free quantum theory of the third-order PU oscillator.

It is an interesting question about the uniqueness of the Hamiltonian function (\ref{H3alt}) for the third-order PU oscillator. Let us suppose that an ansatz
\bea\label{anz}
W=\sum_{n,m=0}^{2}\alpha_{ij}^{nm}x_i^{(n)}x_j^{(m)},
\eea
where $\alpha_{ij}^{nm}=\alpha_{ji}^{mn}$ are arbitrary constants, is the Hamiltonian for the system (\ref{PU3}). Some constants can be expressed in terms of another by imposing that (\ref{anz}) must be an integral of motion. The function $W$ becomes
\bea
W=b_{ij}(\ddot{x}_i+\omega_0^2 x_i)(\ddot{x}_j+\omega_0^2 x_j)+c_{ij}(\dot{x}_i\dot{x}_j-x_i\ddot{x}_j-x_j\ddot{x}_i-\omega_0^2 x_i x_j)+f\epsilon_{ij}\dot{x}_i\ddot{x}_j,
\nonumber
\eea
with $b_{ij}=b_{ji}$, $c_{ij}=c_{ji}$. This function is invariant under the space rotations $\d x_i=\omega_{ij}x_j$, $\omega_{ij}=-\omega_{ji}$ if $b_{ij}=b\d_{ij}$, $c_{ij}=c\d_{ij}$. Thus, we get\footnote{The constants of motion $\ddot{x}_i+\omega_0^2 x_i$ correspond to an invariance of the action (\ref{PU3}) under space translations $\d x_i=a_i$, while conserved charge $\dot{x}_i^2-2 x_i\ddot{x}_i-\omega_0^2 x_i^2$ relates to an invariance under the spatial rotations.}
\bea\label{W}
W=b(\ddot{x}_i+\omega_0^2 x_i)^2+c(\dot{x}_i^2-2 x_i\ddot{x}_i-\omega_0^2 x_i^2)+f\epsilon_{ij}\dot{x}_i\ddot{x}_j.
\nonumber
\eea
If we substitute this function into (\ref{Eq3}), we observe that the corresponding Poisson structure exists only if we put $c=b\omega_0^2$. The redefinition of coefficients $b=\frac{1}{4\omega_0}(\gamma_1+\gamma_2)$, $f=-\frac{1}{2}(\gamma_1-\gamma_2)$ turns to (\ref{H3alt}).

\vskip 0.5cm
\noindent
{\bf{\large 4. The general case}}
\vskip 0.5cm

Let us consider the PU oscillator of the arbitrary odd order with distinct frequencies of oscillation. The equations of motion of this model are
\bea
\prod_{k=0}^{n-1}\left(\frac{d^2}{dt^2}+\omega_k^2\right)\dot{x}_i=\sum_{k=0}^{n}\sigma_{k}^n x_i^{(2k+1)}=0,\;\,\mbox{where}\;\, \sigma_k^n=\sum_{i_1<i_2<..<i_{n-k}=0}^{n-1}\omega_{i_1}^2\omega_{i_2}^2...\omega_{i_{n-k}}^{2},\quad \sigma_{n}^n=1.
\nonumber
\eea
If $H=H(x_i,\dot{x}_i,..,x_i^{(2n)})$ is the Hamiltonian of the $(2n+1)$-order PU oscillator, $\{\cdot,\cdot\}$ is the corresponding Poisson bracket, an analogue of the system of equations (\ref{Eq3}) has a form
\bea\label{HF}
\{x_i^{(m)},H\}=x_i^{(m+1)},\;m=0,1,..,2n-1,\qquad \{x_i^{(2n)},H\}=-\sum_{k=0}^{n-1}\sigma_{k}^n x_i^{(2k+1)}.
\eea
The most evident choice of the Hamiltonian function relates with an integral of motion which corresponds to invariance under time translations. This constant of motion can be presented in a simple form with the aid of the so-called oscillator coordinates which were introduced in paper \cite{Pais}. They are given by
\bea\label{oc}
x_{k,i}=\sqrt{\rho_k}\prod_{m=0\atop m\neq k}^{n-1}\left(\frac{d^2}{dt^2}+\omega_m^{2}\right)x_i=\sqrt{\rho_k}\sum_{m=0}^{n-1}\sigma_{m,k}^{n}x_i^{(2m)},
\eea
where we denote
\bea
\rho_k=\frac{(-1)^k}{\prod\limits_{m=0\atop m\neq k}^{n-1}(\omega_{m}^{2}-\omega_{k}^{2})},\quad
\sigma_{m,k}^{n}=\sum_{i_1<i_2<..<i_{n-m-1}=0\atop i_1,i_2,..,i_{n-m-1}\neq k}^{n-1}\omega_{i_1}^2\omega_{i_2}^2...\omega_{i_{n-m-1}}^{2},\quad \sigma_{n-1,k}^n\equiv 1.
\nonumber
\eea
We choose $0<\omega_0<\omega_1<..<\omega_{n-1}$ for definiteness. The action functional (\ref{odd}) in terms of variables (\ref{oc}) takes a form
\bea\label{actgen}
S=\frac{1}{2}\int\,dt\,\sum_{k=0}^{n-1}(-1)^k\epsilon_{ij}\,x_{k,i}\left(\frac{d^2}{dt^2}+\omega_k^2\right)\dot{x}_{k,j}.
\eea
This representation of the action (\ref{odd}) can be derived by the way which is previously developed for the even-order PU oscillator in the paper \cite{Pais} (for some technical details see also \cite{Masterov_N2PU}). The energy of the model (\ref{actgen}) can be presented as follows
\bea\label{Hamgen}
H=\sum_{k=0}^{n-1}(-1)^{k+1}\epsilon_{ij}\dot{x}_{k,i}\ddot{x}_{k,j}.
\eea
It can be straightforwardly verified that equations (\ref{HF}) are satisfied with respect to this integral of motion. The corresponding Poisson structure relations read
\bea\label{Pois}
\{x_i^{(s)},x_j^{(m)}\}=\left\{
\begin{aligned}
&
0,&& s+m\,-\,\mbox{odd};
\\[2pt]
&
(-1)^{\frac{s-m}{2}+n+1}P_{s+m-2n}(\omega_0^2,\omega_1^2,..,\omega_{n-1}^2)\epsilon_{ij}, && s+m\,-\,\mbox{even},
\end{aligned}
\right.
\eea
where
\bea\label{P}
P_{2k}(\omega_0^2,\omega_1^2,..,\omega_{n-1}^2)=\sum_{\la_{0},\la_1,..,\la_{n-1}=0 \atop\la_0+\la_1+..+\la_{n-1}=k}^{k}\omega_0^{2\la_0}\omega_1^{2\la_1}...
\omega_{n-1}^{2\la_{n-1}}.
\eea
This is the $k$-th degree symmetric polynomial in $n$ variables $\omega_0^2$, $\omega_1^2$,.., $\omega_{n-1}^2$.
By definition, this function is equal to zero for $k<0$.

When verifying the equations (\ref{HF}), the following identities
\bea\label{id1}
\sum_{k=0}^{n-1}(-1)^k\omega_p^{2k}\sigma_{k,s}^{n}=\frac{(-1)^s}{\rho_s}\d_{sp},\quad \sum_{k=0}^{n-1}(-1)^k(-\omega_k^2)^{s}\sigma_{p,k}^{n}\rho_k=
\left\{
\begin{aligned}
&
\d_{sp},&& s=0,1,..,n-1,
\\[5pt]
&
-\sigma_p^n,&& s=n;
\end{aligned}
\right.
\\[2pt]
\label{id2}
P_{2k}(\omega_0^2,\omega_1^2,..,\omega_{n-1}^2)=(-1)^{n-1}\sum_{s=0}^{n-1}(-1)^s \omega_s^{2n+2k-2}\rho_s,\;\mbox{for}\; k=-n+1,-n+2,...,
\eea
prove to be helpful\footnote{One finds the proof of (\ref{id1}) in \cite{Alt}. The proof of identity (\ref{id2}) is given in Appendix A.}.

By analogy with (\ref{coord3}), let us introduce canonical coordinates
\bea\label{canon}
\begin{aligned}
&
q_{k,i}=\sqrt{\frac{1}{2\omega_k}}\left(\dot{x}_{k,1}+\frac{(-1)^i}{\omega_k}\ddot{x}_{k,2}\right),\quad
p_{k,i}=(-1)^k\sqrt{\frac{\omega_k}{2}}\left(\dot{x}_{k,2}+\frac{(-1)^{i+1}}{\omega_k}\ddot{x}_{k,1}\right),
\\[2pt]
&
\qquad\qquad\qquad\qquad z_i=\frac{(-1)^i}{\omega_0\omega_1...\omega_{n-1}}\sum_{k=0}^{n}\sigma_{k}^{n}x_i^{(2k)},\;\mbox{(no sum)},
\end{aligned}
\eea
which obey relations
\bea
\{q_{k,i},p_{m,j}\}=\d_{km}\d_{ij},\qquad \{z_i,z_j\}=\epsilon_{ij},
\nonumber
\eea
with regard to Poisson structure (\ref{Pois}).
The Hamiltonian (\ref{Hamgen}) in terms of these coordinates takes a form
\bea\label{Hgen}
H=\frac{1}{2}\sum_{k=0}^{n-1} (-1)^{k}\left[\left(p_{k,1}^2+\omega_k^2 q_{k,1}^2\right)-\left(p_{k,2}^2+\omega_k^2 q_{k,2}^2\right)\right].
\eea
This representation of the Hamiltonian automatically provides a set of $2n$ positive-definite integrals of motion of the type $J_{k,i}=p_{k,i}^2+\omega_k^2 q_{k,i}^2$ (no sum).
Therefore, let us choose the following ansatz for the alternative Hamiltonian
\bea\label{Hcal}
\mathcal{H}=\frac{1}{2}\sum_{k=0}^{n-1} \left[\gamma_{k,1}\left(p_{k,1}^2+\omega_k^2 q_{k,1}^2\right)+\gamma_{k,2}\left(p_{k,2}^2+\omega_k^2 q_{k,2}^2\right)\right],
\eea
where $\gamma_{k,i}$ are arbitrary nonzero constants. It can be straightforwardly verified that the equations (\ref{HF}) with change $H\rightarrow\mathcal{H}$ are satisfied with regard to the following Poisson structure relations
\bea\label{PB}
\{x_i^{(s)},x_j^{(m)}\}=\left\{
\begin{aligned}
&
0, && s=m=0;
\\[5pt]
&
(-1)^{\frac{s-m+1}{2}}\sum_{k=0}^{n-1}\rho_k\omega_k^{s+m-2}\alpha_k^{+}\d_{ij}, && s+m-odd;
\\[2pt]
&
(-1)^{\frac{s-m}{2}}\sum_{k=0}^{n-1}\rho_k\omega_k^{s+m-2}\alpha_k^{-}\epsilon_{ij}, && s+m\neq 0-even,
\end{aligned}
\right.
\eea
where we denote $\alpha_k^{\pm}=\frac{1}{2}\left(\frac{1}{\gamma_{k,1}}\pm\frac{1}{\gamma_{k,2}}\right)$. It should be noted that this Poisson structure is degenerated when the quantity $s=\sum\limits_{k=0}^{n-1}\frac{\rho_k\alpha_k^{-}}{\omega_k^2}$ vanishes. So, such sets of coefficients $\gamma_{i,k}$ which satisfy relation $s=0$ are discarded from our consideration.

The generalization of coordinates (\ref{qp}), (\ref{pi}) has a form
\bea\label{cf}
\begin{aligned}
&
\mathfrak{q}_{k,i}=\sqrt{|\gamma_{k,i}|}q_{k,i}, \qquad \mathfrak{p}_{k,i}=(-1)^{k+i+1}\mbox{sign}(\gamma_{k,i})\sqrt{|\gamma_{k,i}|}p_{k,i},\quad\mbox{(no sum)},
\\[2pt]
&
\qquad\qquad\qquad\pi_1=\frac{1}{\prod\limits_{r=0}^{n-1}\omega_r\sqrt{|s|}}z_1, \qquad \pi_2=\frac{\mbox{sign}(s)}{\prod\limits_{r=0}^{n-1}\omega_r\sqrt{|s|}}z_2.
\end{aligned}
\eea
These variables obey the following nonvanishing relations
\bea
\{\mathfrak{q}_{k,i},\mathfrak{p}_{m,j}\}=\d_{km}\d_{ij},\qquad \{\pi_i,\pi_j\}=\epsilon_{ij},
\nonumber
\eea
under the Poisson structure (\ref{PB}).

The Hamiltonian (\ref{Hcal}) in terms of coordinates (\ref{cf}) has a form
\bea
\mathcal{H}=\frac{1}{2}\sum_{k=0}^{n-1}\left[\mbox{sign}(\gamma_{k,1})\left(\mathfrak{p}_{k,1}^2+\omega_k^2\mathfrak{q}_{k,1}^2\right)+ \mbox{sign}(\gamma_{k,2})\left(\mathfrak{p}_{k,2}^2+\omega_k^2\mathfrak{q}_{k,2}^2\right)\right].
\nonumber
\eea
If all constants $\gamma_{k,i}$ are positive, this Hamiltonian is positive-definite and more suitable for quantization than (\ref{Hgen}).

It should be noted that the fact of existence of the alternative Hamiltonian formulation for the odd-order PU oscillator is in agreement with the results obtained in paper \cite{Dima_2}.

\vskip 0.5cm
\noindent
{\bf{\large 5. Compatible generalizations}}
\vskip 0.5cm

Let us consider the function
\bea
\tilde{H}=\mathcal{H}+U(x_i,\dot{x}_i,..,x_i^{(2n)}).
\nonumber
\eea
According to the analysis in \cite{Alt}, this function can be viewed as the Hamiltonian of the deformed odd-order PU oscillator whose dynamics is described by
\bea\label{gen}
\sum_{k=0}^{n}\sigma_k^n x_i^{(2k+1)}=\{x_i^{(2n)},U\},
\eea
if the function $U$ obeys the following equations
\bea
\{x_i^{(k)},U\}=0,\quad\mbox{for}\quad k=0,1,..,2n-1,
\nonumber
\eea
with regard to the Poisson structure (\ref{PB}). These equations can be written as follows\footnote{For the case of $n=1$ we put $\rho_0=1$ by definition.}
\bea\label{sys}
\left\{
\begin{aligned}
&
\sum_{k,m=0}^{n-1}(-1)^{m}\rho_k\omega_k^{2p+2m-1}\alpha_k^{+}\frac{\partial U}{\partial x_i^{(2m+1)}}+ \sum_{m=\d_{p,0}}^{n}\sum_{k=0}^{n-1}(-1)^m\rho_k\omega_k^{2p+2m-2}\alpha_k^{-}\epsilon_{ij}\frac{\partial U}{\partial x_j^{(2m)}}=0,
\\[2pt]
&
\sum_{m=0}^{n}\sum_{k=0}^{n-1}(-1)^m\rho_k\omega_k^{2p+2m-1}\alpha_k^{+}\frac{\partial U}{\partial x_i^{(2m)}}- \sum_{k,m=0}^{n-1}(-1)^{m}\rho_k\omega_k^{2p+2m}\alpha_k^{-}\epsilon_{ij}\frac{\partial U}{\partial x_j^{(2m+1)}}=0,
\end{aligned}\right.
\eea
where $p=0,1,..,n-1$. This is the linear homogeneous system of $4n$ equations on $4n+2$ partial derivatives of the function $U$. Consequently, this system has infinitely many solutions. Let us demonstrate it for the case of the third-order PU oscillator. The system (\ref{sys}) for $n=1$ has a form
\bea\label{U1}
\left\{
\begin{aligned}
&
\frac{\alpha_0^{+}}{\omega_0}\frac{\partial U}{\partial\dot{x}_i}-\alpha_0^{-}\epsilon_{ij}\frac{\partial U}{\partial\ddot{x}_j}=0,
\\[2pt]
&
\frac{\alpha_0^{+}}{\omega_0}\frac{\partial U}{\partial x_i}-\alpha_0^{-}\epsilon_{ij}\frac{\partial U}{\partial\dot{x}_j}- \omega_0\alpha_0^{+}\frac{\partial U}{\partial\ddot{x}_i}=0.
\end{aligned}
\right.
\nonumber
\eea
The general solution of this system reads
\bea
U=U\left(\left((\alpha_0^{-})^2-(\alpha_0^{+})^2\right)x_i+\frac{\alpha_0^{-}\alpha_0^{+}}{\omega_0}\epsilon_{ij}\dot{x}_j-
\left(\frac{\alpha_0^{+}}{\omega_0}\right)^2\ddot{x}_i\right).
\nonumber
\eea
This solution takes a form $U=U(x_i)$ if we put $\gamma_{0,1}=-\gamma_{0,2}$.

If the function $U$ is positive-definite, then the deformed odd-order PU oscillator (\ref{gen}) admits the Hamiltonian formulation with regard to positive-definite Hamiltonian. Analogous deformations of the even-order PU oscillators have been originally considered in \cite{KLS,KL} (see also \cite{Alt},\cite{Dima_1})\footnote{About other deformations of the even-order PU oscillator see \cite{Pavsic,Pavsic1} and references therein.}.

\vskip 0.5cm
\noindent
{\large{\bf 6. Conclusion}}
\vskip 0.5cm

To summarize, in this work we have constructed the family of the Hamiltonian structures for the PU oscillator of arbitrary odd order. This was achieved with the aid of several observations. At first, we have shown that the invariance of the third-order PU oscillator under time translations yields the Noether integral of motion which can be presented in the form of the direct sum of two decoupled harmonic oscillators with an alternating sign. Secondly, we have observed that there are oscillator variables (\ref{oc}) whose application to the Lagrangian of the $(2n+1)$-order PU oscillator turns into the direct sum (\ref{actgen}) of $n$ third-order PU oscillators which alternate in a sign.

The alternative canonical formulation for the $(2n+1)$-order PU oscillator, which has been obtained in the present work, corresponds to the Hamiltonian which is a linear combination of $2n$ harmonic oscillators with arbitrary nonzero constants. When all coefficients are positive, we have a positive-definite Hamiltonian which is more viable for physical applications. However, it should be noted that such an oscillator representation of the Hamiltonian does not mean a dynamical equivalence of the $(2n+1)$-order PU oscillator to a system of $2n$ one-dimensional harmonic oscillators. The phase spaces of these systems are not isomorphic. We have also discussed possible deformations of the odd-order PU oscillator which are compatible with the alternative Hamiltonian formulation.

Thus, we apply the approach developed in \cite{Kosinski} to one more set of systems. But some questions related with this method remain actual until now. At first, if a system has a set of integrals of motion which form some Lie algebra under a Poisson bracket corresponding to some conventional approach (e.g. to Dirac's method). Is it possible to deform these integrals of motion in such a way that the same structure relations hold under a Poisson structure which relates with an alternative approach? In other words, can a symmetry structure of a system be reproduced with regard to alternative canonical formalism? Secondly, one of the main motivations of developing alternative ways to obtain Hamiltonian formulations for higher-derivative mechanical systems is their further generalizations to higher-derivative field theories. It is the most argued question about possibility of such generalizations of the alternative method which has been applied in the present work. A generalization of the analysis in this paper to the case of $\,\mathcal{N}=2$ supersymmetric odd-order PU oscillator \cite{Masterov_1} is worth studying as well \cite{FW}.

\vskip 0.5cm
\noindent
{\large{\bf Acknowledgements}}
\vskip 0.5cm
\noindent
The author is grateful to the Institut f\"{u}r Theoretische Physik at Leibniz Universit\"{a}t Hannover for the hospitality. We also thank D. Chow, D. Kaparulin, H. Motohashi, and T. Suyama for useful correspondence. The research was supported by the MSE program Nauka under the project 3.825.2014/K, and RFBR grant 15-52-05022.

\vskip 0.5cm
\noindent
{\large{\bf Appendix.} The proof of identity (\ref{id2})}
\vskip 0.5cm

Let us prove that
\bea
P_{2k}(\omega_0^2,\omega_1^2,..,\omega_{n-1}^2)=(-1)^{n-1}\sum_{s=0}^{n-1}(-1)^s \omega_s^{2n+2k-2}\rho_s,\;\,\mbox{for}\;\,k=-n+1,-n+2,...
\nonumber
\eea
If we take into account the following representation of $\rho_s$
\bea
\rho_s=\frac{1}{V}\left|
\begin{aligned}
&
1 && 1 && .. && 1 && 1 && .. && 1
\\[2pt]
&
\omega_0^2 && \omega_1^2 && .. && \omega_{s-1}^2 && \omega_{s+1}^2 && .. && \omega_{n-1}^2
\\[2pt]
&
\omega_0^4 && \omega_1^4 && .. && \omega_{s-1}^4 && \omega_{s+1}^4 && .. && \omega_{n-1}^4
\\[2pt]
&
.. && .. && .. && .. && .. && .. && ..
\\[2pt]
&
\omega_0^{2n-4} && \omega_1^{2n-4} && .. && \omega_{s-1}^{2n-4} && \omega_{s+1}^{2n-4} && .. && \omega_{n-1}^{2n-4}
\end{aligned}
\right|,
\nonumber
\eea
where $V=\prod\limits_{i_1<i_2=0}^{n-1}(\omega_{i_2}^2-\omega_{i_1}^2)$ is the Vandermonde determinant, then the RHS of (\ref{id2})
can be rewritten in the following equivalent form
\bea
&&
(-1)^{n-1}\sum_{s=0}^{n-1}(-1)^s \omega_s^{2n+2k-2}\rho_s=
\nonumber
\\[2pt]
&&
=\frac{1}{V}\left|
\begin{aligned}
&
1 && 1 && ... && 1
\\[2pt]
&
\omega_0^2 && \omega_1^2 && ... && \omega_{n-1}^2
\\[2pt]
&
... && ... && ... && ...
\\[2pt]
&
\omega_0^{2n-4} && \omega_1^{2n-4} && ... && \omega_{n-1}^{2n-4}
\\[2pt]
&
\omega_0^{2n+2k-2} && \omega_1^{2n+2k-2} && ... && \omega_{n-1}^{2n+2k-2}
\end{aligned}
\right|=
\nonumber
\\[2pt]
&&
=\frac{1}{V}\left|
\begin{aligned}
&
1 && 0 && ... && 0
\\[2pt]
&
\omega_0^2 && \omega_1^2-\omega_0^2 && ... && \omega_{n-1}^2-\omega_0^2
\\[2pt]
&
... && ... && ... && ...
\\[2pt]
&
\omega_0^{2n-4} && \omega_1^{2n-4}-\omega_0^{2n-4} && ... && \omega_{n-1}^{2n-4}-\omega_{0}^{2n-4}
\\[2pt]
&
\omega_0^{2n+2k-2} && \omega_1^{2n+2k-2}-\omega_0^{2n+2k-2} && ... && \omega_{n-1}^{2n+2k-2}-\omega_0^{2n+2k-2}
\end{aligned}
\right|=
\nonumber
\eea
\bea
&&
=\prod_{i=1}^{n-1}(\omega_i^2-\omega_0^2)\frac{1}{V}\left|
\begin{aligned}
&
1 && 1 && ... && 1
\\[2pt]
&
P_2(\omega_0^2,\omega_1^2) && P_2(\omega_0^2,\omega_2^2) && ... && P_2(\omega_0^2,\omega_{n-1}^2)
\\[2pt]
&
P_4(\omega_0^2,\omega_1^2) && P_4(\omega_0^2,\omega_2^2) && ... && P_4(\omega_0^2,\omega_{n-1}^2)
\\[2pt]
&
... && ... && ... && ...
\\[2pt]
&
P_{2n-6}(\omega_0^2,\omega_1^2) && P_{2n-6}(\omega_0^2,\omega_2^2) && ... && P_{2n-6}(\omega_0^2,\omega_{n-1}^2)
\\[2pt]
&
P_{2n+2k-4}(\omega_0^2,\omega_1^2) && P_{2n+2k-4}(\omega_0^2,\omega_2^2) && ... && P_{2n+2k-4}(\omega_0^2,\omega_{n-1}^2)
\end{aligned}
\right|=
\nonumber
\eea
\bea
&&
=\prod_{i_1=1}^{n-1}(\omega_{i_1}^2-\omega_0^2)\prod_{i_2=2}^{n-1}(\omega_{i_2}^2-\omega_1^2)\frac{1}{V}\left|
\begin{aligned}
&
1 && ... && 1
\\[2pt]
&
P_2(\omega_0^2,\omega_1^2,\omega_2^2) && ... && P_2(\omega_0^2,\omega_1^2,\omega_{n-1}^2)
\\[2pt]
&
... && ... && ...
\\[2pt]
&
P_{2n-8}(\omega_0^2,\omega_1^2,\omega_2^2) && ... && P_{2n-8}(\omega_0^2,\omega_1^2,\omega_{n-1}^2)
\\[2pt]
&
P_{2n+2k-6}(\omega_0^2,\omega_1^2,\omega_2^2) &&  ... && P_{2n+2k-6}(\omega_0^2,\omega_1^2,\omega_{n-1}^2)
\end{aligned}
\right|=
\nonumber
\\[2pt]
&&
=...=P_{2k}(\omega_0^2,\omega_1^2,...,\omega_{n-1}^2),
\nonumber
\eea
where we use the identities
\bea
\begin{aligned}
&
\omega_k^{2s}-\omega_m^{2s}=(\omega_k^2-\omega_m^2)P_{2s-2}(\omega_k^2,\omega_m^2),
\\[7pt]
&
P_{2s}(\omega_{i_1}^2,\omega_{i_2}^2,..,\omega_{i_k}^2,\omega_{j_1}^2)-P_{2s}(\omega_{i_1}^2,\omega_{i_2}^2,..,\omega_{i_k}^2,\omega_{j_2}^2)
=(\omega_{j_1}^2-\omega_{j_2}^2)P_{2s-2}(\omega_{i_1}^2,\omega_{i_2}^2,..,\omega_{i_k}^2,\omega_{j_1}^2,\omega_{j_2}^2),
\end{aligned}
\nonumber
\eea
which can be easily proved.

It is evident that $\sum\limits_{s=0}^{n-1}(-1)^s \omega_s^{2n+2k-2}\rho_s=0$ for $k=-n+1,-n+2,..,-1$. This fact is in accordance with the definition (\ref{P}) of the polynomial $P_{2k}(\omega_0^2,\omega_1^2,..,\omega_{n-1}^2)$.

\fontsize{10}{10}\selectfont

\end{document}